\documentclass[aps,prd,twocolumn,fleqn,superscriptaddress]{revtex4}
\usepackage{graphicx,color,natbib}
\usepackage{amsmath,amssymb,amsfonts}

\newcommand{\bse}{\begin{subequations}}
\newcommand{\ese}{\end{subequations}}
\newcommand{\be}{\begin{equation}}
\newcommand{\ee}{\end{equation}}
\newcommand{\bea}{\begin{eqnarray}}
\newcommand{\eea}{\end{eqnarray}}
\newcommand{\ba}{\begin{array}}
\newcommand{\ea}{\end{array}}

\input amssym.def
\input amssym.tex

\usepackage[colorlinks=true, linkcolor=blue, bookmarks=true]{hyperref}

\begin{document}
IPM/P-2015/049

\title{Thermal Quench at Finite t'Hooft Coupling}

\author{M. Ali-Akbari\footnote{$\rm{m}_{-}$aliakbari@sbu.ac.ir}}
\affiliation{Department of Physics, Shahid Beheshti University G.C., Evin, Tehran 19839, Iran}
\author{Hajar Ebrahim\footnote{hebrahim@ut.ac.ir}}
\affiliation{Department of Physics, University of Tehran, North Karegar Ave., Tehran 14395-547, Iran}
\affiliation{School of Physics, Institute for Research in Fundamental Sciences (IPM),
P.O.Box 19395-5531, Tehran, Iran}
\author{ S. Heshmatian\footnote{heshmatian@bzte.ac.ir}}
\affiliation{Department of Engineering Science, Buein Zahra Technical University, Buein Zahra, Qazvin, Iran}

\begin{abstract}
Using holography we have studied thermal electric field quench for infinite and finite t'Hooft coupling constant. The set-up we consider here is D7-brane embedded in ($\alpha'$ corrected) AdS-black hole background. It is well-known that due to a time-dependent electric field on the probe brane, a time-dependent current will be produced and it will finally relax to its equilibrium value. We have studied the effect of different parameters of the system on equilibration time. As the most important results, we have observed a universal behaviour in the rescaled equilibration time in the very fast quench regime for different values of the temperature and $\alpha'$ correction parameter. It seems that in the slow quench regime the system behaves adiabatically. We have also observed that the equilibration time decreases in finite t'Hooft coupling limit.     
\end{abstract}

\maketitle

\tableofcontents

\section{Introduction}
Understanding the properties of a system out-of-equilibrium is a long-standing problem in physics especially when the system is strongly coupled. An experimental example of such systems is the quark-gluon plasma (QGP) produced at RHIC or LHC where two heavy nuclei such as gold or lead are collided at relativistic speeds \cite{CasalderreySolana:2011us}. When the system is strongly coupled usual field theory techniques are not able to describe the properties of the system. Therefore one needs to use other techniques such as lattice gauge theory and AdS/CFT correspondence \cite{CasalderreySolana:2011us, Maldacena}. In this paper we will concentrate on the AdS/CFT approach where the time-dependent systems can be dealt with in real time.   

AdS/CFT correspondence states that ${\cal{N}}=4$ Super-Yang-Milles  (SYM) theory in four dimensions is dual to string theory on $AdS_5 \times S^5$ background. In the most used version of this duality a strongly coupled field theory (infinite t'Hooft coupling constant ($\lambda$) and infinite number of colours ($N_c$)) is dual to classical gravity. In fact the vacuum state in field theory is dual to pure $AdS$ solution and a thermal state to $AdS$-black brane or black hole where the field theory temperature is identified with the black brane or black hole temperature. 

The strongly coupled systems produced in experiments are not infinitely strongly coupled. Therefore it is reasonable to use the gauge/gravity duality in the limit where one studies the effect of the finite t'Hooft coupling constant. In the dual gravity side this corresponds to considering $\alpha '$ corrections which represent the stringy effects in the gravity side \cite{AliAkbari:2010av}. Therefore if one is interested in studying the effect of the finite t'Hooft coupling in the field theory, one needs to deal with the background solution obtained from the gravity action in the presence of $\alpha '$ corrections. Such a solution has been given in \eqref{corrected}. 

An out-of-equilibrium system is produced by the injection of energy in a finite time interval. One way to simulate this situation in gauge/gravity duality is to apply a time-dependent electric field which varies from zero to a finite constant amount \cite{Hashimoto:2013mua, Ali-Akbari:2015gba, Caceres:2014pda}. Therefore the system evolves from an equilibrium state of an initial Hamiltonian to an equilibrium state of the modified Hamiltonian due to the presence of a time-dependent electric field. If the initial state of the system is at non-zero temperature this time-dependent process is usually called thermal quench \cite{Buchel:2014gta}. 

Applying a time-dependent external electric field will produce a time-dependent current in the system which starts from zero and reaches an equilibrium value, corresponding to the final amount of the electric field. This current is the result of interaction between the fundamental degrees of freedom of the system and the electric field. In order to introduce the fundamental matter in the AdS/CFT picture, we have to add probe branes to the background dual to the strongly coupled system under study \cite{Karch:2003nh}. 
 
In this paper we are interested in studying the effect of temperature and finite t'Hooft coupling on the equilibration in a strongly coupled system. The observable that can be examined to see how the system equilibrates due to the time-dependent external electric field is the behaviour of time-dependent current produced in the system \cite{Ali-Akbari:2015gba}. This time-dependent current will reach its equilibrium value after some time which we call it equilibration time. We will see how this equilibration time modifies with the change in the parameters of the system.

\section{Time-Dependent External Electric Field}

Here we consider a general class of black hole metrics of the form 
\be\label{metric}%
ds^2 = G_{tt} dt^2 +G_{xx} d{\vec{x}}^2+G_{zz} dz^2+ G_{ss}d\Omega_5^2 ,
\ee%
which is $AdS_5\times S^5$, asymptotically. $z$ is the radial coordinate and the boundary of the above background is located at $z=0$. The four-dimensional spacetime coordinates where the field theory lives are denoted by $t$ and $\vec{x}$. The five-dimensional sphere is shown by $d\Omega_5^2$ and its metric can be written as %
\be %
d\Omega_5^2= d\theta^2+\cos^2\theta d\Omega_3^2+\sin^2\theta d\varphi^2 .
\ee %
In order to add the fundamental matter to the field theory, it is well known that D-branes must be added to the background in the probe limit which means that the branes do not back-react on the background. In this limit the background metric \eqref{metric} is fixed and therefore, according to AdS/CFT correspondence, the dynamics of the fundamental matter in the field theory is explained by the Dirac-Born-Infeld (DBI) action. For a D7-brane, the DBI action is given by 
\begin{eqnarray}
S_{\rm D7} = - \mu_7  \int\! dt d^3\vec{x} dz d\Omega_3 \,e^{-\phi} \sqrt{-\det 
\left[
g_{ab} + 2\pi \alpha' F_{ab}
\right]}\, ,
\end{eqnarray}
where the static gauge, meaning that the eight directions on the brane have been identified with $(t, \vec{x}, z, \Omega_3)$, has been applied. Upon static gauge fixing, $a, b\ \epsilon\ (t, \vec{x}, z, \Omega_3)$. $F_{ab}$ is the field strength of the gauge field living on the brane and induced metric on the brane $g_{ab}$ is defined as 
\be %
 g_{ab}=G_{MN}\partial_a X^M \partial_b X^N ,
\ee %
where $M, N\ \epsilon\ (t, \vec{x}, z, \Omega_5)$. $\phi$ is Dilaton field and it is non-trivial in the cases we are studying in the following. 

To have a time-dependent electric field along one of the field theory directions, say $x$, we need to consider   $A_x(t,z)$ to be non-zero. We have also assumed that $\varphi = 0$ and $\theta = 0$ which means we are dealing with massless fundamental degrees of freedom. As it is clear the $x$-component of the gauge field is not a function of $\vec{x}$ since we would like to consider a homogeneous electric field quench. Therefore the Lagrangian gets the following general form
\be\begin{split} %
{\cal{L}}&\propto G_{xx} G_{ss}^{3/2}\,e^{-\phi}\cr
&\times \sqrt{(2\pi\alpha')^2G_{tt}F_{zx}^2 - G_{zz}\left((2\pi\alpha')^2F_{tx}^2-G_{tt}G_{xx}\right)} ,\cr
&= G_{xx} G_{ss}^{3/2}\sqrt{\chi} ,
\end{split}\ee %
and the resulting equation of motion for $A_x(t,z)$ becomes
\be%
\partial_z\left(\frac{\,e^{-\phi}\, G_{ss}^{3/2} G_{xx} G_{tt}F_{zx}}{\sqrt{\chi}}\right)-\partial_t\left(\frac{\,e^{-\phi}\,G_{ss}^{3/2} G_{xx} G_{zz}F_{tx}}{\sqrt{\chi}}\right)=0 .
\ee%
Now an appropriate ansatz for the gauge field to cause a time-dependent electric field in the $x$ direction is \cite{Hashimoto:2013mua} 
\begin{eqnarray}
\label{Eansatz}
A_x=-\int^tE(s)ds+a(t,z) \, ,
\end{eqnarray} 
where $E(t)$ is the time-dependent electric field which injects energy into the system. $E(t)$ can be chosen to have different forms which describes various ways of energy injection. It should be emphasized that, at least, a number of these different choices lead to the same final equilibrium state of the system\cite{Ali-Akbari:2015gba}. The equation of motion for $A_x$ leads to a second order partial differential equation for $a(t,z)$. Similar to the near boundary expansion in the static external electric field case \cite{Karch:2007pd},  the time-dependent current in the field theory is given by the second derivative of $a$ with respect  to $z$ at the boundary 
\be%
\label{current}
j(t) \propto \partial_z^2 a(t,z=0).
\ee%

\section{Energy Injection and Equilibration  }
Following the discussion in the previous section we choose the form of the time-dependent electric field to be
\be%
E(t)=\frac{E_0}{2}\left(1+\tanh(\frac{t}{k})\right).
\ee%
The electric field is zero at infinite past and constant value $E_0$ at infinite future. The transition time, $k$, is the period of time during which the electric field reaches from zero to its final value $E_0$. Using the ansatz \eqref{Eansatz} the equation of motion for the gauge field, coming from the DBI action, leads to a second order nonlinear equation in both $t$ and $z$ for the field $a(t,z)$. In order to solve this equation one needs to impose two boundary conditions and two initial conditions. The boundary conditions are $a(t,z) = \partial_z a(t,z) = 0$ on the boundary and the initial conditions are $a(t_0,z) = \partial_t a(t_0,z) = 0$ at some initial time $t_0$. After solving this equation, one obtains the time-dependent current using \eqref{current}. At large time the time-dependent current approaches its static value that is \cite{Karch:2007pd}
\be%
j_{st}= \left(e^{-\phi}\sqrt{G_{tt}} G_{xx} G_{ss}^{3/2}\right)_{z=z_*},
\ee%
where $z_*$ can be found from 
\be %
G_{tt} G_{xx}-(2\pi\alpha')^2 E^2=0.
\ee %
One can define an equilibration time as $\epsilon (t_{eq}) < 0.05$ where 
\be%
\epsilon (t) =\frac{j(t) - j_{st}}{j_{st}}.
\ee%
According to this equation we define the equilibration time as the time where the time-dependent current approaches the static one with 5\% uncertainty and stays in this regime afterwards. In this section we will evaluate the equilibration time for two different backgrounds and see how it modifies with the relevant parameters. 

\subsection{Infinite t'Hooft Coupling Constant }
The background we would like to study is the black brane solution
\be\begin{split}%
-G_{tt} &= z^{-2} (1-w^4) ,\cr
G_{xx} &= z^{-2} ,\cr
G_{zz} &= z^{-2} (1-w^4)^{-1} ,\cr 
G_{ss} &= 1 ,
\end{split}\ee%
where $w=\frac{z}{z_h}$ and $z_h$ is the horizon. According to the AdS/CFT correspondence the above black brane solution is dual to a thermal state in SYM theory at infinite t'Hooft Coupling with infinite number of colours. The black brane temperature $T=\frac{1}{\pi z_h}$ is identified with the dual field theory temperature. 

\subsection{Finite t'Hooft Coupling Constant}
In this subsection we concentrate on the following asymptotically AdS solution \cite{AliAkbari:2010av}
\be\label{corrected}\begin{split}%
-G_{tt} &= z^{-2} (1-w^4) T(w) ,\cr
G_{xx} &= z^{-2} X(w) ,\cr
G_{zz} &= z^{-2} (1-w^4)^{-1} U(w) ,\cr 
G_{ss} &= 1+2 S(w) ,
\end{split}\ee%
where 
\be\begin{split} %
T(w) &= 1-b (75 w^4+\frac{1225}{16} w^8-\frac{695}{16} w^{12}) + {\cal{O}} (b^2) ,\cr
X(w) &= 1-\frac{25 b}{16} w^8(1+w^4) + {\cal{O}} (b^2) ,\cr
U(w) &= 1+b (75 w^4+\frac{1175}{16} w^8-\frac{4585}{16} w^{12}) + {\cal{O}} (b^2) ,\cr
S(w) &= \frac{15 b}{32} w^8(1+w^4) + {\cal{O}} (b^2) ,\cr
\phi(w) &= -\ln g_s - \frac{45 b}{8} (w^4+\frac{1}{2} w^8 + \frac{1}{3} w^{12}) + {\cal{O}} (b^2) ,
\end{split}\ee%
where $b=\frac{\zeta (3)}{8} \lambda^{-3/2}$, $g_s$ is the string coupling constant and as in the previous subsection $w=\frac{z}{z_h}$. This metric is the solution to the bulk equations of motion in the presence of the $\alpha'$ corrections. As mentioned in the introduction it corresponds to having finite t'Hooft coupling corrections in the field theory side. It should be noted that the temperature of the above solution is $T=\frac{1}{\pi(1-b)z_h}$.

\section{Result}
We will discuss the results for infinite and finite t'Hooft coupling in two following subsections.
\begin{figure}
\centering
\includegraphics[width=70mm]{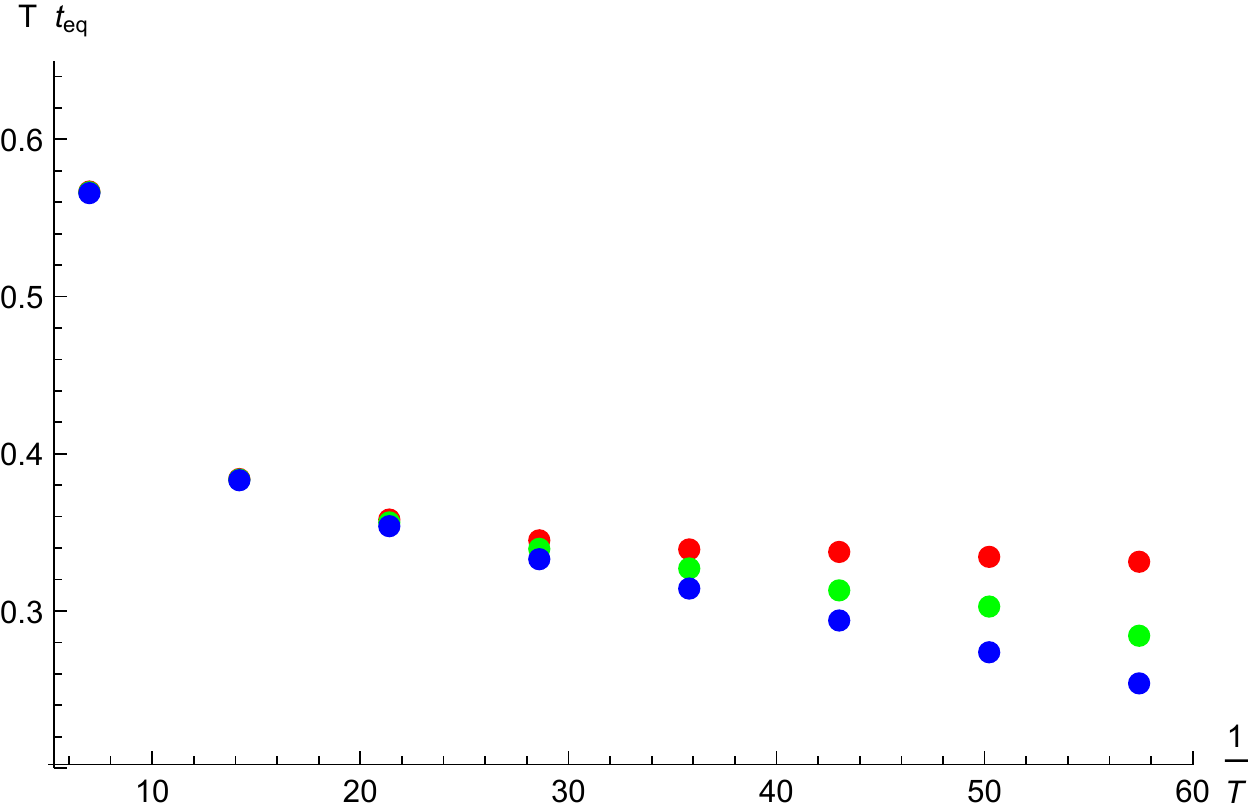}
\caption{$T t_{eq}$ has been plotted with respect to inverse temperature for $k^{-1}=0.7$ and $b = 0$. Red, green and blue points correspond to $E0 = 
 0.001,~0.004$ and $0.006$, respectively. The rescaled value of the equilibration time falls with decreasing the temperature.}\label{TtoneoverTb0}
\end{figure}

\begin{figure}
\centering
\includegraphics[width=70mm]{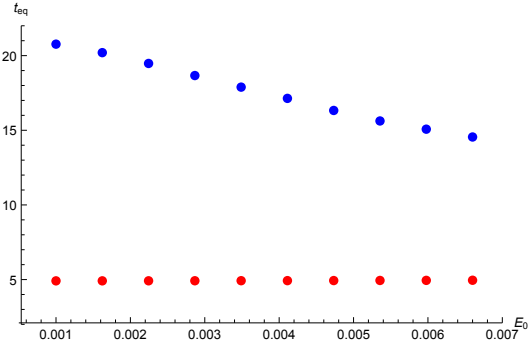}
\caption{$t_{eq}$ with respect to the final value of the electric field is plotted for $k^{-1} =0 .7$ and $b = 0$. Red (Blue) points represent the equilibration time for a system at $T=0.14~(0.01)$. At higher temperatures of the system, the change in the equilibration time with raising the value of the final electric field is less noticeable. } \label{tEb0}
\end{figure} 

\begin{figure}
\centering
\includegraphics[width=70mm]{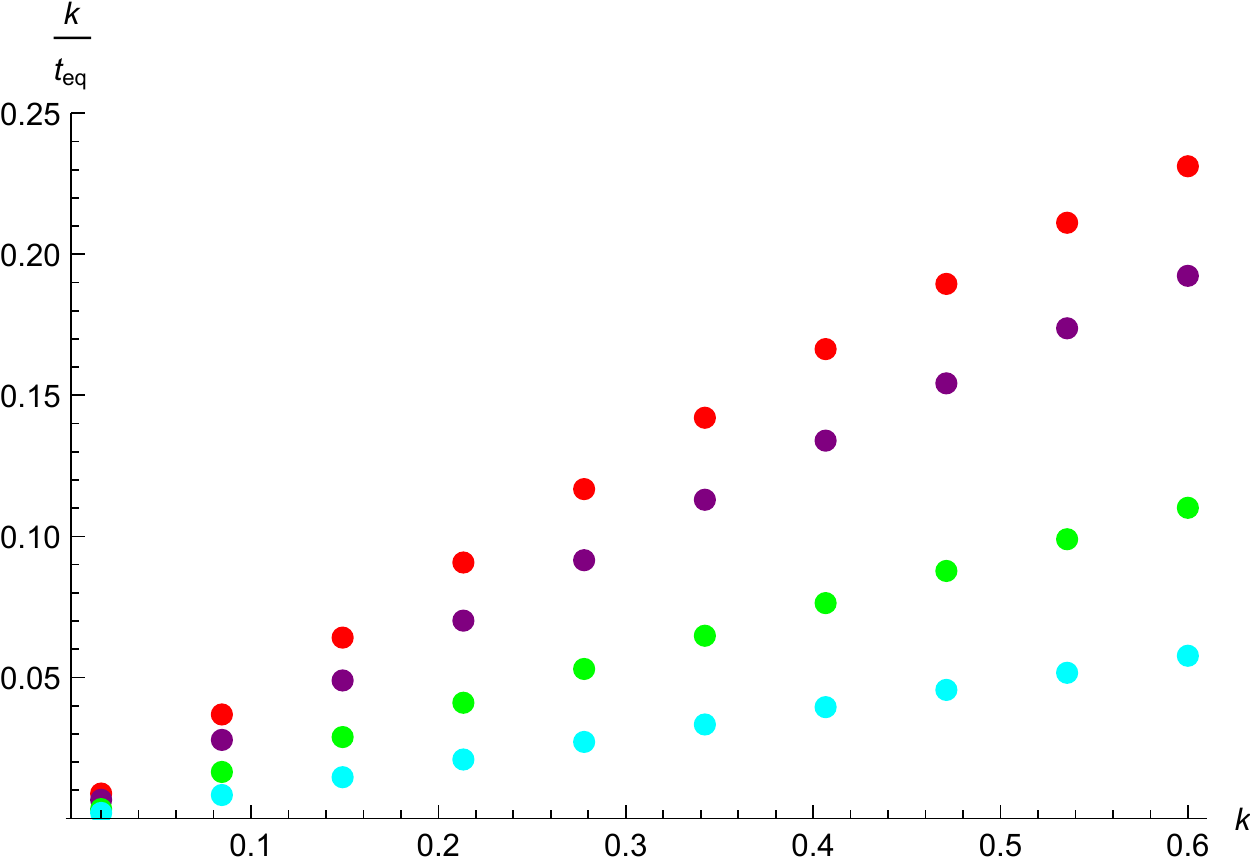}
\caption{This figure shows $k t_{eq}^{-1}$ with respect to $k$ for $E_0=0.001$ and $b = 0$. The values of $k$ have been chosen to lie in the fast quench regime. The points in red, purple, green and cyan colours represent $T = 0.14,~0.1,~0.06$ and $0.03$, respectively. A universal behaviour by which we mean the independence of $k t_{eq}^{-1}$ of the system temperature is observed in the fast quench regime.} \label{kovertkb0fast}
\end{figure}

\begin{figure}
\centering
\includegraphics[width=70mm]{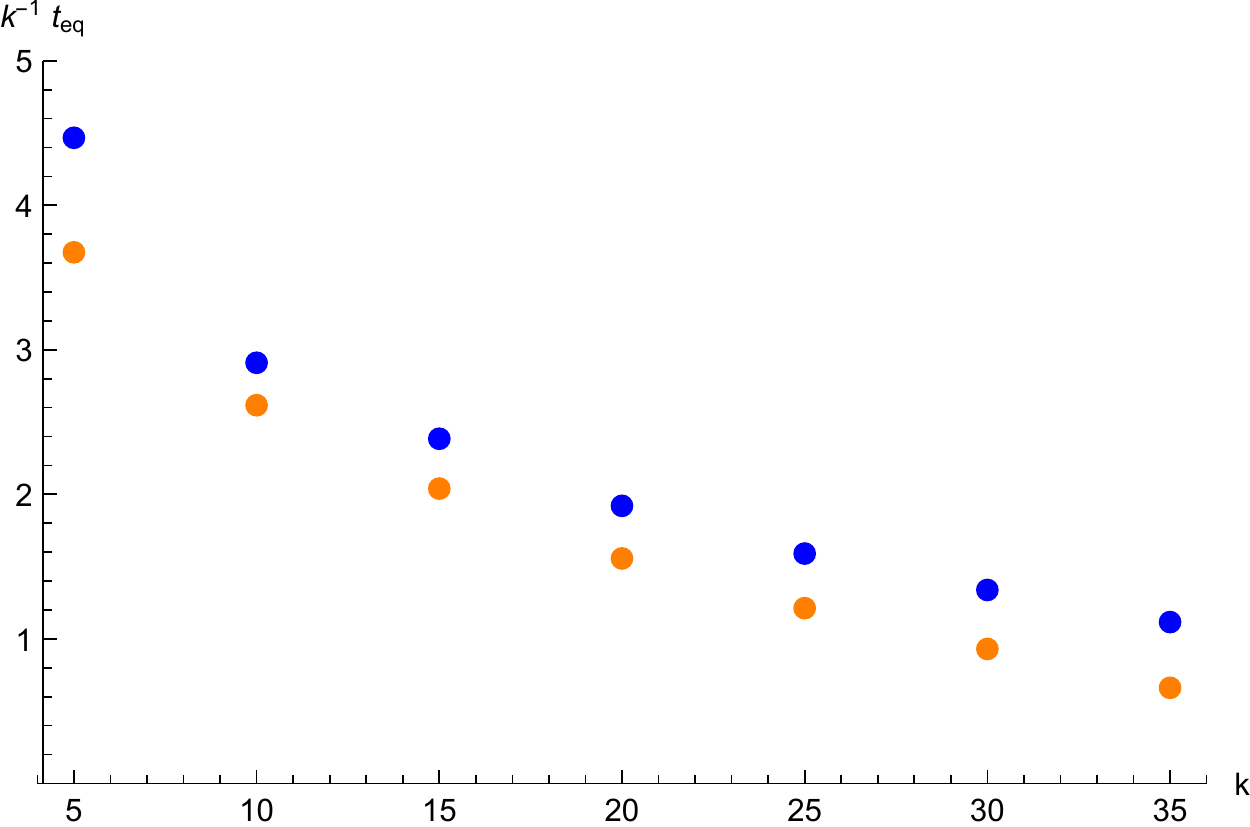}
\caption{In the slow quench regime $k^{-1} t_{eq}$ with respect to $k$ has been plotted for $E_0=0.001$ and $b=0$. The orange and blue points correspond to $T = 0.02$ and $T = 0.01$, respectively. It seems that the system shows an adiabatic behaviour in this regime.} \label{toverkkb0slow}
\end{figure}

\begin{figure}
\centering
\includegraphics[width=70mm]{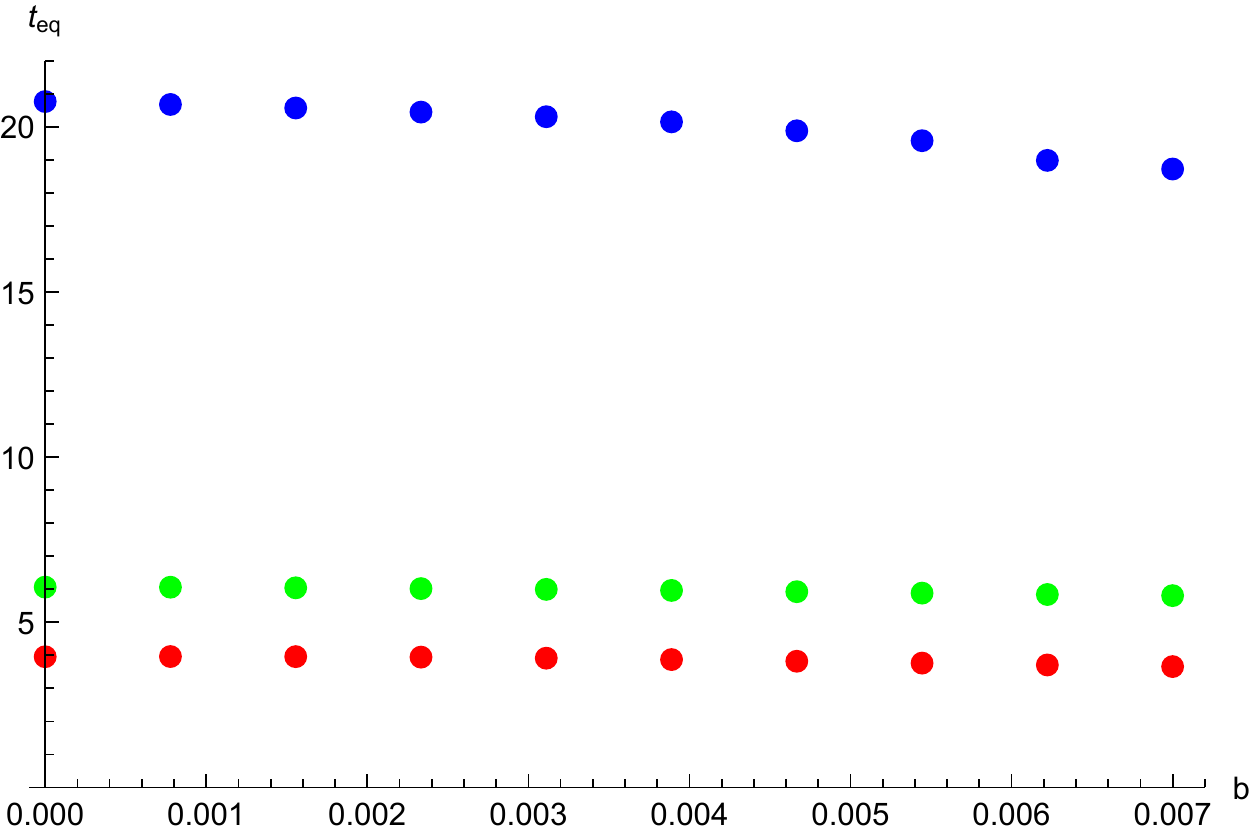}
\caption{At finite t'Hooft coupling we have plotted the equilibration time with respect to the correction parameter $b$ for $k^{-1}=0.7$ and $E_0 = 0.001$. The red, green and blue points present different temperatures of the system, $T = 0.14,~0.06$ and $0.01$, respectively. The equilibration time has a mild dependence on $b$, especially for higher values of the temperature.}\label{tb}
\end{figure}

\begin{figure}
\centering
\includegraphics[width=70mm]{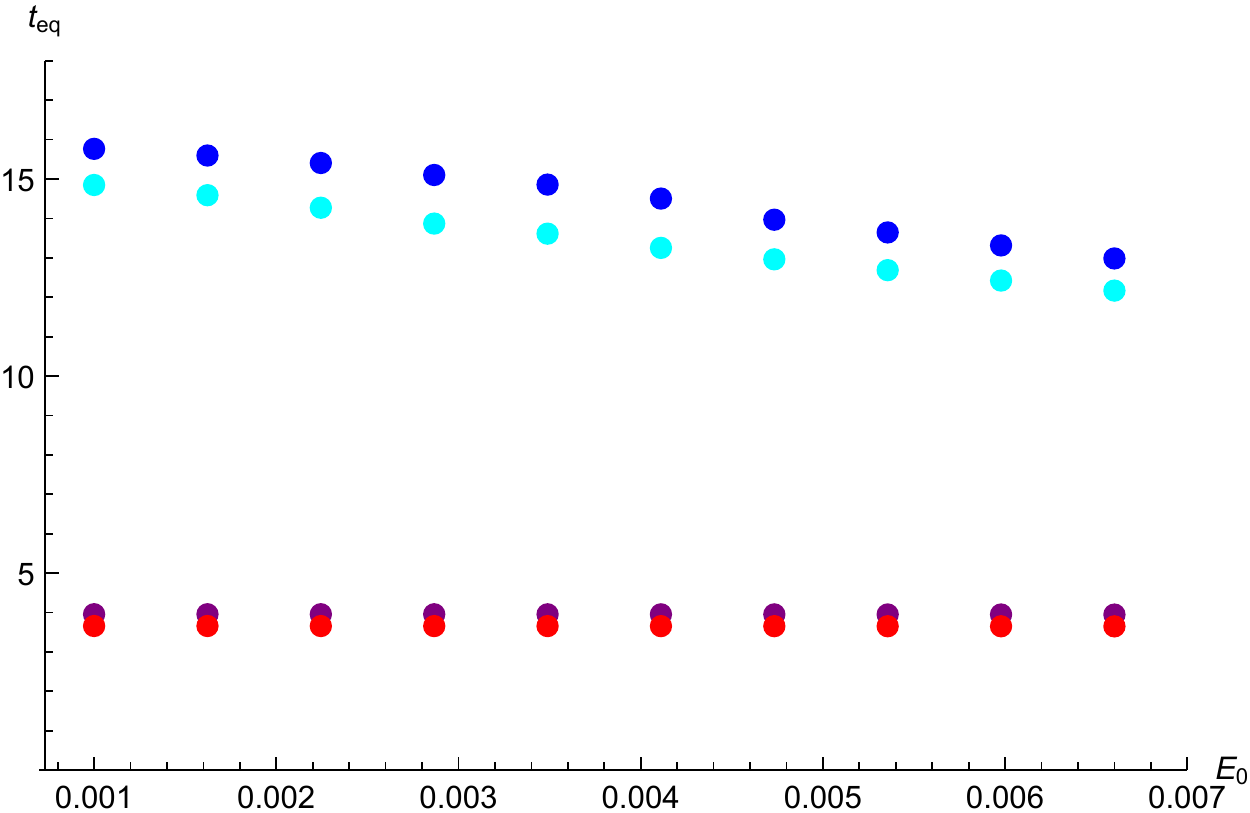}
\caption{$t_{eq}$ versus $E_0$ for $k^{-1}=0.7$ is plotted. We have compared infinite ($b=0$) and finite t'Hooft coupling ($b=0.007$) results. For $T = 0.02$, blue (cyan) points show $b=0~(0.007)$ and for $T = 0.14$, purple (red) points present $b=0~(0.007)$. The effect of corrections is more noticeable at lower temperatures.}\label{tEb}
\end{figure}

\begin{figure}
\centering
\includegraphics[width=70mm]{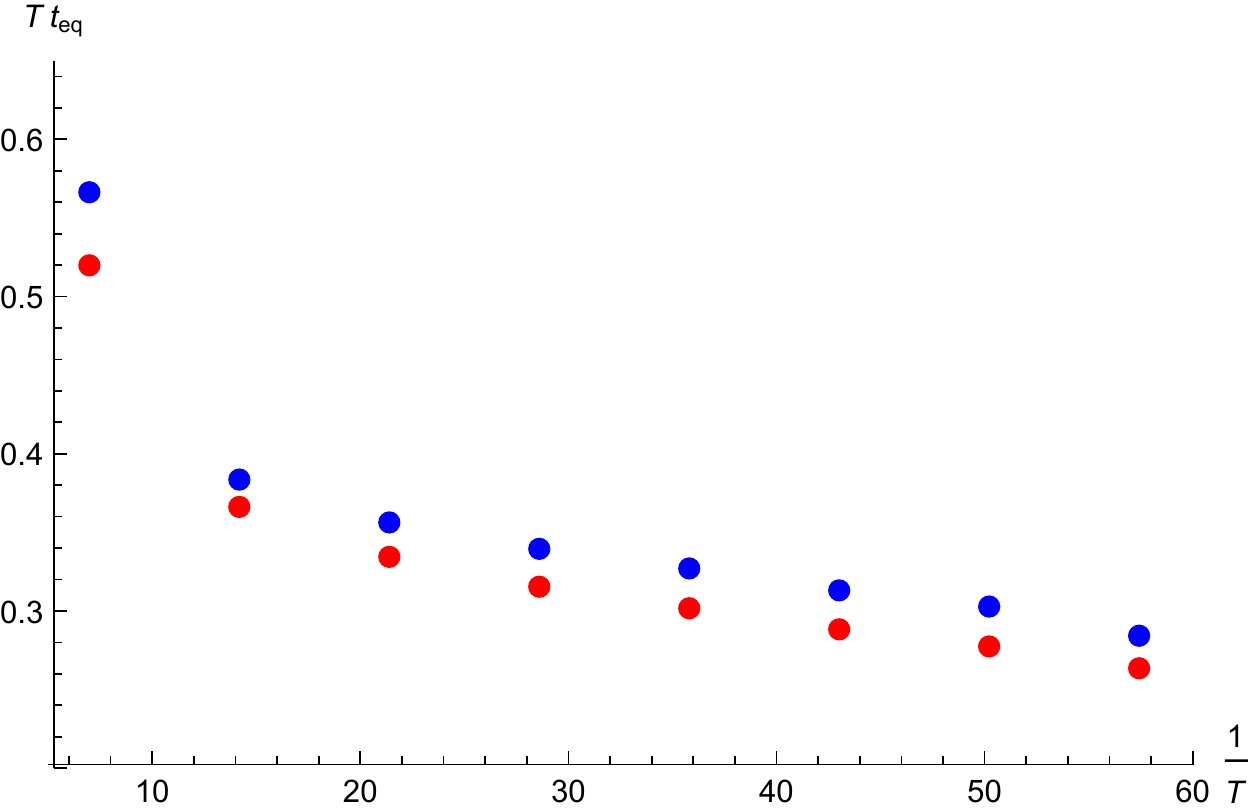}
\caption{$T t_{eq}$ with respect to $T^{-1}$ for $k^{-1}=0.7$ and $E_0=0.004$  where blue (red) points correspond to $b=0~(0.007)$. $\alpha'$ correction decreases the rescaled equilibration time as the temperature is reduced.}\label{TtoneoverTb}
\end{figure}

\begin{figure}
\centering
\includegraphics[width=70mm]{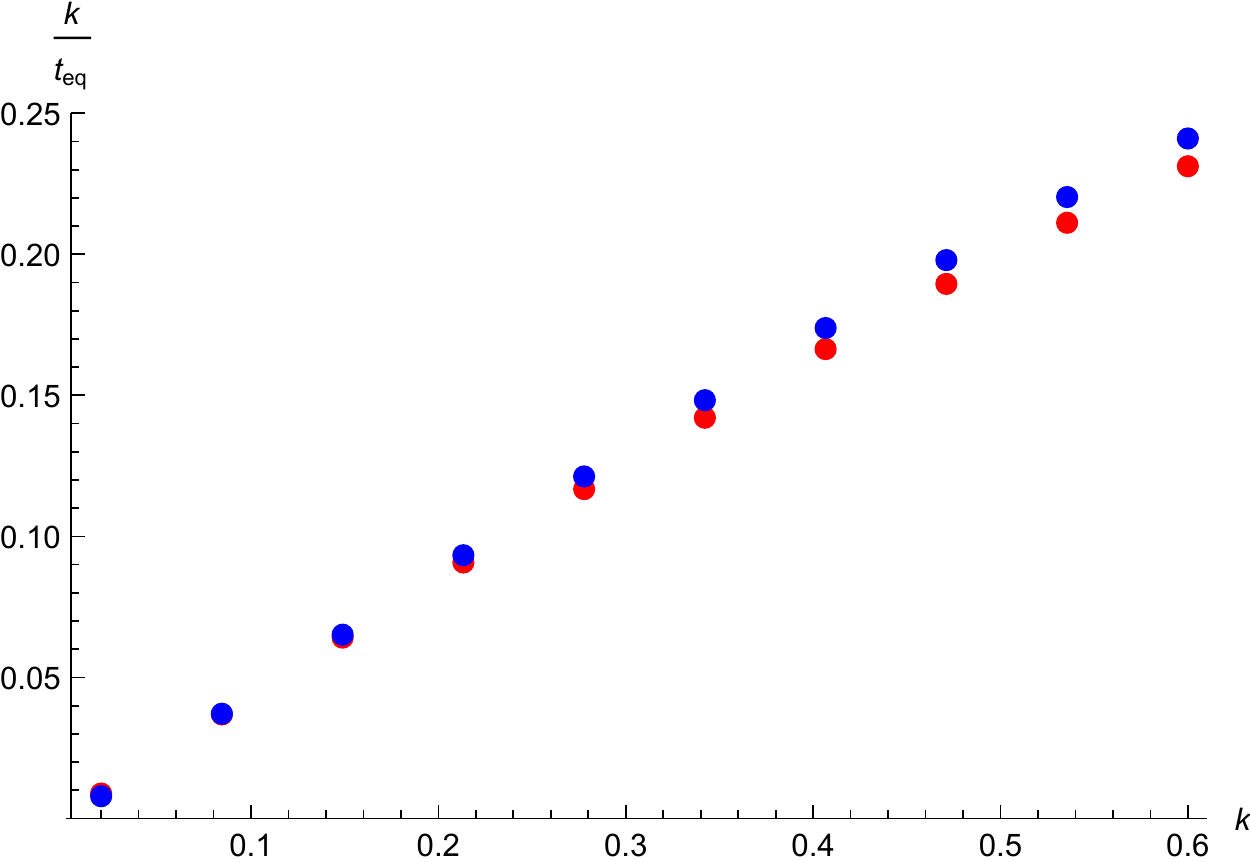}
\caption{The plot shows $k t_{eq}^{-1}$ versus $k$ for $E_0=0.001$ and $T = 0.14$ in the fast quench regime where red (blue) points correspond to $b=0~(0.007)$. At very small values of $k$ the points for different couplings merge.} \label{kovertkbfast}
\end{figure}

\begin{figure}
\centering
\includegraphics[width=70mm]{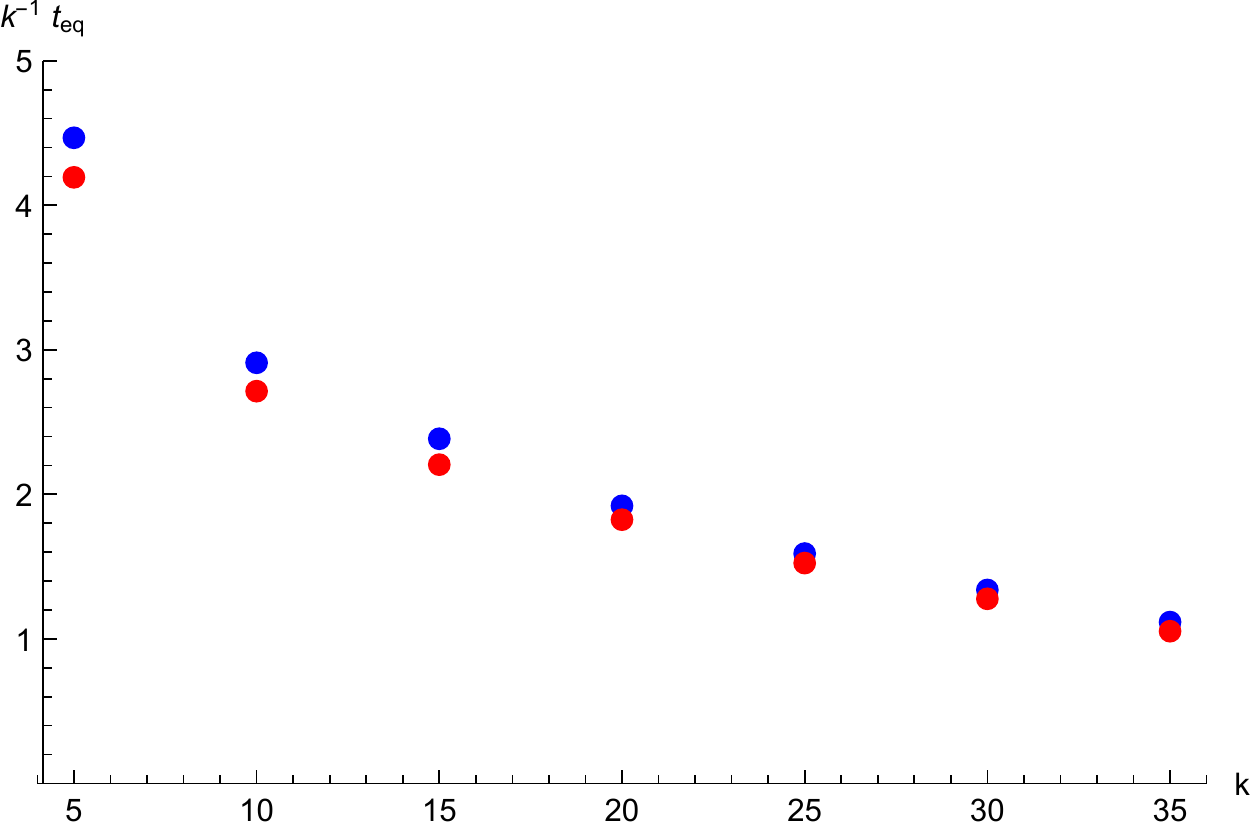}
\caption{$k^{-1} t_{eq}$ with respect to $k$ is plotted for $E_0=.001$ and $T = 0.01$ in the slow quench regime. The blue (red) points demonstrate $b=0~(0.007)$. Again we will see that at very large values of $k$, adiabatic regime, the points coincide.} \label{toverkkbslow}
\end{figure}

\subsection{Infinite t'Hooft Coupling}
We begin with $b=0$ results where the effect of finite t'Hooft coupling is ignored. We have plotted the equilibration time with respect to different parameters in the problem. Figure \ref{TtoneoverTb0} shows the dependence of $T t_{eq}$ on inverse temperature for various values of the electric field. One can define a relevant time-scale for the theory as $\frac{1}{T}$ which gives the time-scale at which the perturbation around the equilibrium configuration in the system relaxes.  Figure \ref{TtoneoverTb0} shows that the equilibration time with respect to this time-scale decreases as the temperature of the system is reduced. This seems reasonable since as the temperature is raised the thermal fluctuations do not allow the current produced by the applied time-dependent electric field reach its static value quickly. This is confirmed by the plots for different values of the final electric field. At low temperatures where the thermal fluctuations are small, the larger value of the final electric field the less equilibration time. But at high temperature the equilibration time does not change by varying the final value of the electric field. This means that, for the range of the electric field considered in this paper, at higher temperatures the response of the system is dominated by the thermal fluctuation effect.

In figure \ref{tEb0} we have plotted the equilibration time with respect to the final value of the electric field for two different fixed temperatures. One observes that for higher temperature (red points) the sensitivity of the system to the time-dependent electric field (in the range we have considered in this paper) is ignorable compared to the lower value of the temperature (blue points). This confirms the last conclusion we reached in figure \ref{TtoneoverTb0}. On an equal footing,  the blue points in figure \ref{tEb0} show that at low temperature the equilibration time decreases as the value of the final electric field is raised. So that the system equilibrates faster as we expected.  

In the above results we had assumed that the transition time $k$ is kept fixed. One of the interesting things that can be discussed in this context is to examine the regimes of slow and fast quenches. These are two different limits of the time-scale during which the electric field changes to reach a constant value. This has been studied in the literature extensively such as \cite{Ali-Akbari:2015gba, Buchel:2014gta, Buchel:2013gba}. The electric field varies slowly in a long period of time during the slow quench. This corresponds to having $k\gg 1$. While the fast quench is the opposite limit, $k\ll 1$. In figure \ref{kovertkb0fast}, fast quench regime, we have plotted $k t_{eq}^{-1}$ with respect to $k$ for different values of the temperature while the final value of the electric field is kept the same. We can observe that at larger values of $k$, longer periods of electric field time-dependent change, the points corresponding to different temperatures are widely separated and as $k$ is reduced the points start approaching a single value. This behaviour shows that for very fast quenches ($k\ll 1$) the value of $k t_{eq}^{-1}$ is independent of temperature and this indicates a universal behaviour. Such universal behaviour means that the equilibration time is the same for systems at different temperatures if the electric field changes abruptly. Similar behaviour occurs for various values of the electric field while the temperature of the system is the same, as has been also observed in \cite{Ali-Akbari:2015gba}. 

In the opposite regime where $k\gg 1$, slow quench, $k t_{eq}^{-1}$ decreases as $k$ is raised, figure \ref{toverkkb0slow}. It shows that for the longer time-scales of the electric field change the system has enough time to adjust with the energy injection and equilibrate. Thus we can speculate that at $k \rightarrow \infty$ the system passes through equilibrium points in its phase space which, in fact, describes the adiabatic behaviour.  

\subsection{Finite t'Hooft Coupling}
In this subsection we will see how the finite coupling or in other words the $\alpha '$ corrections will affect the result we discussed previously. In figure \ref{tb} the dependence of equilibration time on the correction parameter $b$ for different values of the temperature has been shown. Note that the transition time $k$ and the final value of the electric field is kept fixed. Interestingly we observe that at high enough temperatures (red and green points) the system is not sensitive to the change of the parameter $b$. However at lower temperatures (blue points) the dependence of the equilibration time on $b$ is more significant. In fact increasing $b$ leads to a decrease in the equilibration time.

The next figure, \ref{tEb}, where we have plotted the equilibration time with respect to $E_0$ for fixed $k$,  also confirms the conclusion made in the previous paragraph. Different colours show different values of $b$ and $T$. The effect of the $\alpha '$ corrections on the equilibration time is more significant at lower temperature (blue and cyan points) and cause the equilibration time to decrease. It seems, as these figures confirm, the finite t'Hooft coupling corrections affect the system similarly to the enhancement of  electric field. 

We have also studied the effect of the corrections on the $T t_{eq}$ with respect to inverse temperature, figure \ref{TtoneoverTb}, for one of the temperatures discussed in figure \ref{TtoneoverTb0}. We again see that in the presence of the finite t'Hooft coupling the equilibration time decreases in the $\frac{1}{T}$ scale. We have also obtained the effect of the finite t'Hooft coupling on $k t_{eq}^{-1}$ and $k^{-1} t_{eq}$ in the fast (figure \ref{kovertkbfast}) and slow (figure \ref{toverkkbslow}) quench regimes, respectively. In both cases the rescaled equilibration time decreases compared to the infinite t'Hooft coupling results. A worthy observation is that in the very fast and adiabatic regimes the points presenting infinite and finite t'Hooft coupling coincide and the effect of $\alpha'$ corrections is negligible. It seems that at these two regimes the behaviour of the system, even in the presence of the finite t'Hooft coupling corrections, is described by infinitely coupled SYM theory.

\end{document}